\documentclass[a4paper,10pt]{article}
\usepackage{graphicx} 

\title{Anomalous Diffusion of particles with inertia in external potentials}
\author{S.~Eule, R.~Friedrich and F.~Jenko}

\begin{document}

\maketitle

\begin{abstract}
Recently a new type of Kramers-Fokker-Planck Equation has been proposed [R. Friedrich et al. Phys. Rev. Lett. {\bf 96}, 230601 (2006)]  describing anomalous diffusion
in external potentials. In the present paper the explicit cases of a harmonic potential and a velocity-dependend damping are incorporated. Exact relations for moments for these cases are presented and the asymptotic behaviour for long times is discussed. Interestingly the bounding potential and the additional damping by itself lead to a subdiffussive behaviour, while acting together the particle becomes localized for long times.
\end{abstract}

\section{Introduction}
In recent years the concept of Continous Time Random Walks (CTRWs) \cite{Montroll} has been extensively used to model
systems that exhibit the phenomenon which has been termed "Anomalous Diffusion". Various mechanisms in physical systems are known leading to sub- and superdiffusion respectively. CTRWs are thus suitable to model
a variety of physical, chemical, biological and even medical systems, like charge transport in 
disordered systems \cite{Scher}, protein dynamics \cite{Glockle}, transport in low-dimensional chaotic systems \cite{Zasla}, anomalous transport in plasmas\cite{Castillo, Sanchez, Balescu} and the spread
of a pandemic \cite{Brockmann}. For a review see e.g. \cite{Metzler}. CTRWs in a natural way lead to a description within the framework of  fractional kinetic equations and in particular Fractional-Fokker-Planck Equations (FFPEs).\\
In a recent publication \cite{Eule}, \cite{CTRW} the concept of CTRWs has been extended to particles with finite inertia under the influence of a stochastic force. In the context of this model the particle is subjected to a series of random kicks, such that the particle's motion is deterministic most of the time. This leads to a generalized equation of Kramers-Fokker-Planck (KFP) type which is characterized by a collision operator that is nonlocal in space as well as in time. For a special choice of the time-evolution kernel a Fractional Kramers-Fokker-Planck Equation has been obtained. Fractional kinetic equations in general and fractional diffusion equations, in particular, have a wide range of applications. We should mention that over the past years several generalized KFP-equations \cite{BaSil}, \cite{Metzlerklafter}, \cite{Sokolov1} have been proposed (for a review including discussions of various applications see e.g., Ref.~\cite{Coffey}). In a recent paper \cite{Langevin} the connection of these equations is established unsing an approach based on the underlying Langevin equations. \\
In this paper we want to extend the discussions of the force-free case started in \cite{CTRW} and discuss the effects of an inclusion of a linear force and an additional damping between the kicks. This case should be of importance for various practical applications in the context of anomalous diffusion in random and accordingly complex environments.\\
For results concerning anomalous transport of particles without inertia in external fields the reader is refered to
\cite{MKS}. Recent results on CTRWs in an oscillating external field are given in \cite{Sokolov}.\\
This paper is organized as follows. First we present the model under consideration and incorporate the effects of
the harmonic potential and the damping. As we will see, the resulting equation will be rather complicated.
We thus concentrate on the equations for the moments and discuss in this context some limiting cases. We will present analytical and numerical results for the time-evolution of the second-order moments.

\section{The model}

Some 70 years ago Kramers investigated the motion of Brownian particles in a fluid.
In a seminal paper \cite{Kramers} he considered the joint-probability-distribution of position and velocity $f({\bf x, u}, t)$ for which he could derive the well-known equation:
\begin{equation}\label{Kramers}
   \left[{\partial\over\partial t}+\nabla_x\cdot{\bf u}
   +\nabla_u \cdot {\bf A}({\bf x,u})\right]
   f({\bf x},{\bf u},t) = {\cal L}_{\rm FP}f({\bf x},{\bf u},t)
\end{equation}
where ${\bf A}({\bf x,u})$ is the acceleration due to an external potential and ${\cal L}_{\rm FP}$ is the Fokker-Planck collision operator \cite{Risken}, \cite{vanKampen}
\begin{equation}\label{FPop}
   {\cal L}_{\rm FP}f=\gamma\nabla_u\cdot({\bf u}f)+K\,\Delta_u f\,.
\end{equation}
Note that this notation indicates that the divergences act on the product of the following function and the probability distribution.\\
In \cite{Eule} a new type of the Kramers-Fokker-Planck Equation has been proposed. In the context of this model the motion of the particle is deterministic while from time to time it is subjected to a kick in a random direction with random velocity. As the time $\tau$ between the kicks as well as the new velocity ${\bf u}$ is given by some probability distribution, a CTRW-model in the velocity-space has been obtained.
As a main result from the following master equation
\begin{eqnarray}
 & & \left[{\partial\over\partial t}+\nabla_x\cdot{\bf u}
   +\nabla_u \cdot {\bf A}({\bf x,u})\right]f({\bf x},{\bf u},t) = \nonumber \\
 & & \int_0^t dt'\,\Phi(t-t')\,\int d{\bf u'}F({\bf u};{\bf u'}) {\cal P}^{t,t'}f({\bf x},{\bf u'},t')\nonumber \\
 & & -\int_0^t dt'\,\Phi(t-t')\, {\cal P}^{t,t'}f({\bf x},{\bf u},t')
\end{eqnarray}
the generalized Kramers equation
\begin{eqnarray}\label{central}
   & & \left[{\partial\over\partial t}+\nabla_x\cdot{\bf u}
   +\nabla_u \cdot {\bf A}({\bf x,u})\right]f({\bf x},{\bf u},t) = \nonumber \\
   & & = {\cal L}_{\rm FP}\int_0^t dt'\,\Phi(t-t')\,
   {\cal P}^{t,t'}f({\bf x},{\bf u},t')\, ,
\end{eqnarray}
has been derived, where $F({\bf u};{\bf u'})d{\bf u'}$ denotes the probability that the particle's velocity
will end up in the velocity space element $d{\bf u}$ about ${\bf u}$ and $\Phi(t-t')$ is some memory kernel. The deterministic propagation between the random kicks is 
governed by the Frobenius-Perron operator
\begin{eqnarray}
  & & {\cal P}^{t,t'}f({\bf x},{\bf u},t') = e^{-(t-t')[\nabla_x\cdot{\bf u}+
  \nabla_u \cdot {\bf A}({\bf x,u})]}\,f({\bf x},{\bf u},t') \, . 
\end{eqnarray}
Eq. (\ref{central}) generalizes Eq. (\ref{Kramers}) in two aspects. First it incorporates effects which are
nonlocal in time and second it contains retardation effects due to ${\cal P}^{t,t'}$ which renders the
equation nonlocal in space as well. For the special case of an asymptotic power-law waiting-time distribution 
$\Psi ( \tau ) \sim \tau ^{-(1+\delta)}$ one obtains $\int_0^t\Phi(t-t') \sim \,_0{\mathcal D}_t^{1-\delta}$ where $_0{\mathcal D}_t^{1-\delta}$ is the Riemann-Liouville fractional derivative. For an excellent account on CTRWs the reader is referred to \cite{Balescu}. A detailed derivation of Eq. (\ref{central}) with an explicit definition of 
$F({\bf u};{\bf u'})$ is found in \cite{Eule}.

\section{Damped motion in a harmonic potential}
In this section we shall consider the special case, where the external potential is given by 
${\bf A(x,u)}=-2\eta {\bf u}-\omega_0^2 {\bf x}$. The corresponding equations of motion for an individual particle
without the impact of fluctuations read:
\begin{equation}\label{dynd}
   \dot{\bf x}(t)={\bf u}(t)\,,\quad \dot{\bf u}(t)={\bf A}({\bf x,u})= -2\eta {\bf u}-\omega_0^2 {\bf x}\,.
\end{equation}
This set of equations determines the behavior of an individual particle between two succesive kicks. As we have 
already mentioned the effect of the Perron-Frobenius-Operator is to project the position and velocity of the particle back to the last kick. For this reason we have  to solve Eq. (\ref{dynd}) and invert this solution to retain the retardation
of the probability distribution function (pdf) on the right hand side of Eq. (\ref{central}). The solution reads
\begin{eqnarray}\label{equom}
{\bf x}(t)& = & e^{-\eta t}\left({\bf x_0} \cos(\omega'(t-t'))+\frac{{\bf u_0}+\eta {\bf x_0}}{\omega '}\sin (\omega '(t-t'))\right) \nonumber \\
{\bf u}(t)& = & e^{-\eta t}\left(\frac{-\eta{\bf u_0}-(\eta ^2+{\omega '}^2 ){\bf x_o}}{\omega '}\sin(\omega' (t-t')+{\bf u_0} \cos(\omega'(t-t')\right) \, .
\end{eqnarray}
Hereby we imposed the initial conditions ${\bf x}(t')={\bf x_0}$, ${\bf u}(t')={\bf u_0}$ and used the abbrevation
$\omega ' =\sqrt{\omega_0^2-\eta^2}$. Note that for $\eta > \omega_0$, $\omega '$ becomes purely imaginary and we
have an exponential relaxation of ${\bf x}(t)$ and ${\bf u}(t)$. Inverting this equations gives a rather complicated expression and solving Eq. (\ref{central}) with this retardation would be a hopeless task. In the remainder of this paper we thus first restrict ourselves to the special cases where we have either  set $\omega_0^2=0$ or one of the damping constants $\eta=0$, $\gamma=0$ respectively. For these special cases rigorous results are presented in the next sections. 

\section{Damped motion between the kicks}

\begin{figure}
\begin{center}
\includegraphics[height=6cm]{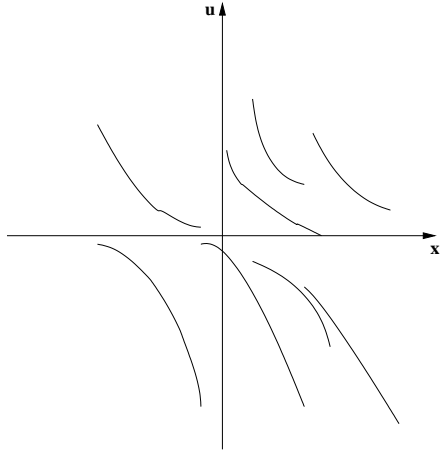}
\caption{Visualization of damped motion between the kicks in phase-space}\label{damp}
\end{center} 
\end{figure}

In this section the limiting case $\omega_0\ll \eta$ is studied i.e. we will analyze purely damped motion between
the kicks. The motion of such a particle in phase-space is visualized in figure \ref{damp}. The motion of an individual particle between the kicks is given by
\begin{equation}\label{dynd1}
   \dot{\bf x}(t)={\bf u}(t)\,,\quad \dot{\bf u}(t)={\bf A}({\bf u})= -2\eta {\bf u}\, ,
\end{equation}
with the solution
\begin{eqnarray}\label{eqom1}
{\bf x}(t)&=&{\bf x_0}+{\bf{u_0}}\frac{1}{2\eta}\left(1-e^{-2\eta(t-t')}\right)\nonumber \\
{\bf u}(t)&=&{\bf u_0}e^{-2\eta(t-t')}\, .
\end{eqnarray}
As already mentioned we have to invert Eq. (\ref{eqom1}) to get the retardation of the pdf in Eq. (\ref{central}), i.e. we have to regard the effect of the Perron-Frobenius-operator
\begin{eqnarray}\label{eqom1inv}
{\bf x_0}&=&{\bf x}(t)-{\bf u}(t)\frac{1}{2\eta}\left(1-e^{2\eta(t-t')}\right) \nonumber \\
{\bf u_0}&=&{\bf u}(t)e^{2\eta(t-t')}\, .
\end{eqnarray}
To calculate the temporal behavior of the moments we use Eq. (\ref{central}). Here we have to be careful. On the lefthand side of (\ref{central}) we have to average over the values at $t$ while on the righthand side we average over retarded values at $t'$. Consequently we get retardation effects in the moment equations. Assuming the system to be infinite and spatially homogeneous we obtain for the moments of lowest order 
\begin{equation}
\frac{\partial}{\partial t}{\bf q}(t)={\bf A}{\bf q}(t)+\int_0^t\Phi(t-t')\left({\bf B}{\bf q}(t') +{\bf I}\right)\,dt' \, . 
\end{equation}
Here we have introduced
\begin{equation}
{\bf q}=\left(
\begin{array}{c}
\langle{\bf x^2}\rangle \\
\langle{\bf ux}\rangle \\
\langle{\bf u^2}\rangle
\end{array}\right), \, 
{\bf I}=\left(
\begin{array}{c}
0\\
0\\
2K
\end{array}\right), \,
{\bf A}=\left(
\begin{array}{c c c}
0 & 2 & 0\\
0 & -2\eta & 1\\
0 & 0 & -4\eta
\end{array}\right)
\quad\mbox{and}
\end{equation}
\begin{equation}
{\bf B}=-\gamma\left(
\begin{array}{c c c}
0 & 0 & 0\\
0 & e^{-2\eta(t-t')} & e^{-2\eta(t-t')}\left(1-e^{-2\eta(t-t')}\right)\\
0 & 0 & -2e^{-4\eta(t-t')}
\end{array}\right)\, .
\end{equation}

It is straightforward to generalize these equations for moments of any order.\\
Due to the occurence of the convolution integrals it is appropriate to switch to the Laplace domain,
where this set of equations can be solved.
\begin{equation}\label{u2lap}
\langle{\bf u}^2\rangle(\lambda)=\frac{2K\Phi(\lambda)}{\lambda+4\eta+2\gamma\Phi(2\eta+\lambda)} \, ,
\end{equation}
\begin{equation}\label{uxlap}
\langle{\bf u}{\bf x}\rangle(\lambda)=\frac{1-\frac{\gamma}{2\eta}\Phi(2\eta+\lambda)+\frac{\gamma}{2\eta}\Phi(4\eta+\lambda)}{\lambda+2\eta+
\gamma\Phi
(2\eta+\lambda)}\langle{\bf u}^2\rangle(\lambda)
\end{equation}
\begin{equation}\label{x2lap}
\langle{\bf x}^2\rangle(\lambda)=\frac2\lambda \langle{\bf u}{\bf x}\rangle(\lambda)\, .
\end{equation}
Hereby we have used the Laplace shifting theorem. It is apparent how the additional damping affects the 
time evolution kernel. Obviously the asymptotic behaviour for large times , i.e. the small $\lambda$-behaviour, is given by the Laplaceinversion of $\Phi(\lambda)$ for $\langle{\bf u^2}\rangle(\lambda)$ and $\langle{\bf ux}\rangle(\lambda)$ and by $\frac{\Phi(\lambda)}{\lambda}$ for $\langle{\bf x^2}\rangle(\lambda)$ respectively.\\
To obtain concrete results we consider the important case of a fractional evolution kernel, i.e. $\Phi(\lambda)=\lambda^{1-\delta}$. We get
\begin{equation}
\langle{\bf u}^2\rangle\sim\frac{K}{2\eta +\gamma(2\eta)^{1-\delta}}\lambda^{-\delta}
\end{equation}
which corresponds to the asymptotic behaviour
\begin{equation}
\langle{\bf u}^2\rangle(t)\sim\frac{K}{\Gamma(\-\delta)(2\eta +\gamma(2\eta)^{1-\delta})}t^{1-\delta}\, ,
\end{equation}
\begin{equation}
\langle{\bf ux}\rangle(t)\sim\frac{\gamma((2\eta)^{1-\delta}+(4\eta)^{1-\delta})}{4\eta ^2+\gamma (2\eta)^{2-\delta}}\frac{K}{\Gamma(\delta)(2\eta +\gamma(2\eta)^{1-\delta})}t^{1-\delta}\, .
\end{equation}
For the mean-square-displacement we obtain
\begin{equation}\label{msdeta0}
\langle{\bf x}^2\rangle(t)\sim\frac{2\gamma((2\eta)^{1-\delta}+(4\eta)^{1-\delta})}{4\eta ^2+\gamma (2\eta)^{2-\delta}}\frac{K}{\Gamma(1+\delta)(2\eta +\gamma(2\eta)^{1-\delta})}t^{\delta}\, .
\end{equation}
The additional damping thus shifts the subballistic-superdiffusive behaviour to the subdiffusive regime. From Eqs. (\ref{u2lap})-(\ref{x2lap}) it is clear that the magnitude of the damping constant is crucial for the rate of convergence to the asymptotics. In the classical limit, i.e. $\delta=1$ standard diffusive behaviour $\langle{\bf x}^2\rangle(t)\sim t$ is obtained.\\
To complete our discussions we mention that for $\eta\to 0$ we get the results described in \cite{CTRW}.
For $\gamma\to 0$ we obtain
\begin{equation}
\langle{\bf u}^2\rangle\sim \frac{K}{\Gamma(\delta)\, 2\eta}t^{1-\delta}\, ,
\end{equation}
\begin{equation}
\langle{\bf ux}\rangle \sim \frac{K}{\Gamma(\delta)\, 4\eta^2}t^{1-\delta}\, ,
\end{equation}
and
\begin{equation}
\langle{\bf x^2}\rangle\sim \frac{K}{\Gamma(1+\delta) \, 4\eta^2}t^{\delta}\, .
\end{equation}
We see that even the additional damping alone leads to subdiffusive behaviour. This is in contrast
to the free flight case, where even if $\gamma\neq 0$ superdiffusive-suballistic behaviour is obtained. 

\section{Response to an external harmonic potential}

\begin{figure}
\begin{center}
\includegraphics[height=6cm]{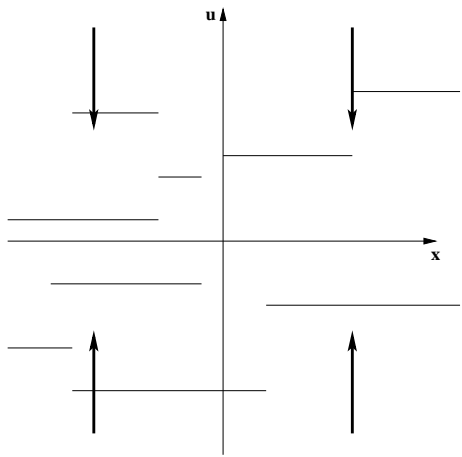}
\caption{The effect of $\gamma$ in Eq. (\ref{central}) visualized for free motion between the kicks}\label{gammavis}
\end{center} 
\end{figure}

\begin{figure}
\begin{center}
\includegraphics[height=6cm]{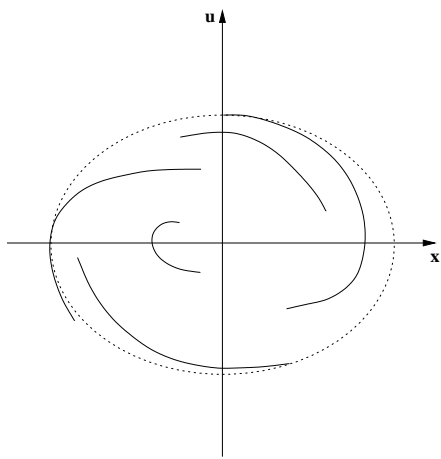}
\caption{The combined effect of the harmonic potential and damping in phase-space}\label{dampedho}
\end{center} 
\end{figure}

\begin{figure}
\begin{center}
\includegraphics[height=4cm]{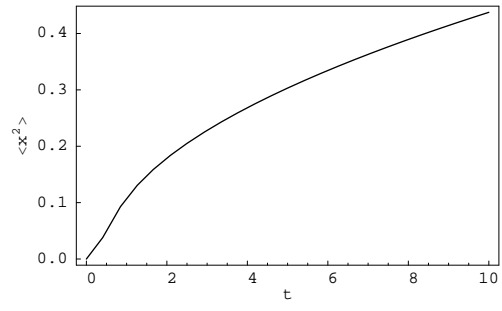}
\caption{$\omega_0=0$, $\eta=2$}\label{omega0bild}
\end{center} 
\end{figure}

\begin{figure}
\begin{center}
\includegraphics[height=4cm]{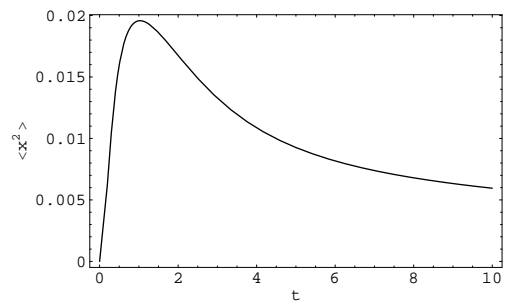}
\caption{Overdamped motion between the kicks, $\eta=4$, $\omega_0=2$}\label{overdamped}
\end{center} 
\end{figure}

Let us now examine the motion of the particles exposed to an external linear restoring force. In other words we will study the effects of an external harmonic potential acting on the particles between the kicks. As mentioned before discussing this problem with both dampings acting on the particle leads to bulky equations. For this reason we consider first the situation when $\eta\ll \omega_0^2$. Eq. (\ref{equom}) then reads
\begin{eqnarray}\label{eqomho}
{\bf x}(t) &=& {\bf x_0} \cos (\omega(t-t'))+\frac{{\bf u_0}}{\omega}\sin (\omega(t-t'))\nonumber \\
{\bf u}(t) &=& -\omega{\bf x_0} \sin (\omega(t-t'))+{\bf u_0}\cos (\omega(t-t'))\, ,
\end{eqnarray}
where we have imposed the initial conditions ${\bf x}(t')={\bf x_0}$ and ${\bf u}(t')={\bf u_0}$.
In this context a word of caution shall be appropriate. One has to be careful in setting the initial conditions.\\
We can now repeat the procedure of the previous chapter for the calculation of the moments. The non-locality in space provides for a mixing of the space and velocity coordinates.\\
We obtain
\begin{equation}\label{hodynsys}
\frac{\partial}{\partial t}{\bf q}(t)={\bf A}{\bf q}(t)+\int_0^t\Phi(t-t')\left({\bf B}{\bf q}(t') +{\bf I}\right)\,dt' \, . 
\end{equation}
Here we have introduced
\begin{equation}
{\bf q}=\left(
\begin{array}{c}
\langle{\bf x^2}\rangle \\
\langle{\bf ux}\rangle \\
\langle{\bf u^2}\rangle
\end{array}\right), \, 
{\bf I}=\left(
\begin{array}{c}
0\\
0\\
2K
\end{array}\right), \,
{\bf A}=\left(
\begin{array}{c c c}
0 & 2 & 0\\
-\omega_0 & 0 & 1\\
0 & -2\omega_0 & 0
\end{array}\right)
\qquad \mbox{and} \qquad
\end{equation}
\begin{equation}
{\bf B}=-\gamma\left(
\begin{array}{c c c}
0 & 0 & 0\\
-\frac{\omega_0^2}{2}\sin (2\omega_0(t-t'))& \cos(2\omega_0(t-t')) & \frac{1}{2\omega_0}\sin (2\omega_0(t-t'))\\
2\omega_0^2 \sin^2(\omega_0(t-t')) & -2\omega_0 \sin(2\omega_0(t-t')) & \cos^2(\omega_0(t-t'))
\end{array}\right)\, .
\end{equation}
Switching to Laplace space we obtain a linear algebraic system
\begin{equation}
\lambda\, {\bf q}(\lambda)= {\bf C}(\lambda){\bf q}(\lambda)+{\bf I}
\end{equation}
where
\begin{equation}
{\bf C}=\left(
\begin{array}{c c c}
0 & 2 & 0\\
\frac{\gamma \omega_0}{4i}(A-B)-\omega_0 & -\frac{\gamma}{2}(A+B) & -\frac{\gamma}{4i\omega_0}(A-B)+1\\
\frac{\gamma \omega_0^2}{2}(A+B)-\gamma\omega_0^2 \Phi(\lambda) & \frac{\gamma}{i}(A-B)-2\omega_0^2 & -\frac{\gamma}{2}
(A+B)+\gamma \Phi(\lambda)
\end{array}\right)
\end{equation}
and ${\bf I}=\left(0, 0, 2K\frac{\Phi(\lambda)}{\lambda}\right)$. Hereby we have used the short-hand notation $A=\Phi(\lambda-2i\omega_0)$ and $B=\Phi(\lambda+2i\omega_0)$. Note that the
relationship $\Phi(z^*)=\Phi^*(z)$ holds and thus $A^*=B$, where the $(\cdot)^*$ means complex conjugation. It follows
in this case that ${\bf C}$ has only real-valued entries.\\
In the Laplace-domain we obtain a linear system of equations which is solvable but we have not yet been able to transform these expressions back or to extract useful limiting cases. One should notice that for this system the knowledge of just the long-time behaviour is no longer sufficient. In particular the oscillations of the moments for small $t$ are ignored. Employing the fractional time kernel we get for $\delta=1$ the ordinary Kramers equation. It is interesting to note how in this case the additional couplings in the dynamical system (\ref{hodynsys}) disappear.\\
\begin{figure}
\begin{center}
\includegraphics[height=6cm]{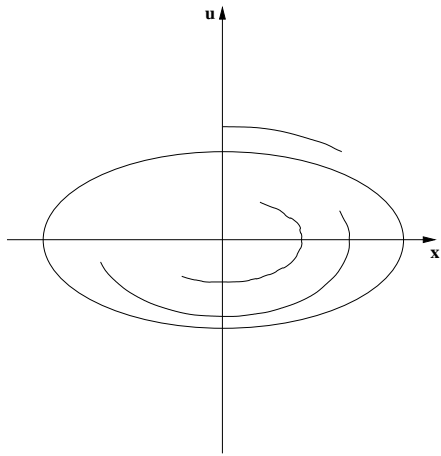}
\caption{Visualization of the motion in a harmonic potential}\label{freeho}
\end{center} 
\end{figure}
Let us now concentrate on the general case for $\gamma =0$. The equations for the moments up to second order are then no longer different from a retardation free version of Eq. (\ref{central}) (i.e. the following results for the moments stated here also hold for the Kramers equation proposed by Barkai and Silbey \cite{BaSil}). This allows us to obtain concrete results on the behaviour of the moments.\\ 
It is important to first consider the physical meaning of $\gamma$ in  Eq.(\ref{FPop}). In the classical Kramers equation the damping operates between two successive random events. In the generalized Kramers equation Eq.(\ref{central}) the damping acts during the kicks. A positive $\gamma$ affects the probability $F({\bf u}, {\bf u'})$ in a way that it is more probable to have smaller absolute value of ${\bf u'}$ after a kick. This effect is visualized in figure \ref{gammavis}.
Without this damping the energy of the particle would tend to infinity for
the force-free case. If we have an additional damping between the kicks in the generalized Kramers eq. though we can set $\gamma=0$ without loss of physical significance. Thus the limiting case $\gamma \to 0$ is not solely of academic interest. The phase-portrait of such a motion is shown in figure \ref{dampedho}.\\
The set of equations for the second order moments reads in Laplace space
\begin{equation}
\langle{\bf u^2}\rangle(\lambda)=\frac{2K \Phi(\lambda)(\lambda^2+2\eta \lambda+2\omega_0^2)}{\lambda(2\eta+\lambda)(\lambda^2
+4\eta\lambda+4\omega_0^2)}\, ,
\end{equation}
\begin{equation}
\langle{\bf ux}\rangle(\lambda)=\frac{2K \Phi(\lambda)}{(2\eta+\lambda)(\lambda^2
+4\eta\lambda+4\omega_0^2)}\, ,
\end{equation}
\begin{equation}
\langle{\bf x^2}\rangle(\lambda)=\frac{4K \Phi(\lambda)}{\lambda(2\eta+\lambda)(\lambda^2
+4\eta\lambda+4\omega_0^2)}\, .
\end{equation}
To obtain concrete results we consider once again the case $\Phi(\lambda)=\lambda^{1-\delta}$
\begin{equation}
\langle{\bf u^2}\rangle(\lambda)=\frac{2K \lambda^{-\delta}(\lambda^2+2\eta \lambda+2\omega_0^2)}{(2\eta+\lambda)(\lambda^2
+4\eta\lambda+4\omega_0^2)}\, ,
\end{equation}
\begin{equation}
\langle{\bf ux}\rangle(\lambda)=\frac{2K \lambda^{1-\delta}}{(2\eta+\lambda)(\lambda^2
+4\eta\lambda+4\omega_0^2)}\, ,
\end{equation}
\begin{equation}
\langle{\bf x^2}\rangle(\lambda)=\frac{4K \lambda^{-\delta}}{(2\eta+\lambda)(\lambda^2
+4\eta\lambda+4\omega_0^2)}\, .
\end{equation}
The inverse Laplace-Transforms of these expressions can now be explicitly calculated.\\
For the mean-square-displacement (MSD) $\langle{\bf x^2}\rangle(t)$ we obtain
\begin{eqnarray}
\langle{\bf x^2}\rangle(t)=&\frac{1}{\hat{\omega}\Gamma(\delta)}\left(2^{-\delta-1}e^{-2t(\eta+\sqrt{\hat{\omega}}})Kt^\delta  
\left(2 e^{2t\sqrt{\hat{\omega}}}(\Gamma(\delta, -2\eta t)-\Gamma(\delta))(\eta t)^{-\delta}+e^{4t\sqrt{\hat{\omega}}}\right.\right.\\ \nonumber &(t(\sqrt{\hat{\omega}}-\eta))^{-\delta}(\Gamma(\delta)-\Gamma(\delta,(2t\sqrt{\hat{\omega}}-\eta))) 
\left.\left.+(-t(\sqrt{\hat{\omega}}-\eta))^{-\delta}(\Gamma(\delta)-\Gamma(\delta,(-2t\sqrt{\hat{\omega}}-\eta)))\right)\right)
\end{eqnarray}
Hereby ${\hat \omega}$ is given by ${\hat \omega}=\eta^2-\omega_0^2$ and $\Gamma(\alpha,\beta)$ is the incomplete
Gamma-function. (The occurence of these incomplete Gamma-functions, which are integral expressions, is due to the multiple occurence of convolution-integrals when backtransforming the moment-equations)\\
We do not want to state the other complete expressions here but rather discuss some qualitative features of the
results. Let us in the following focus our attention on the mean-square-displacement. We present numerical results for some interesting limiting cases and discuss these results. The following results were obtained for $\delta=0.5$ and $K=1$.\\
\begin{figure}
\begin{center}
\includegraphics[height=4cm]{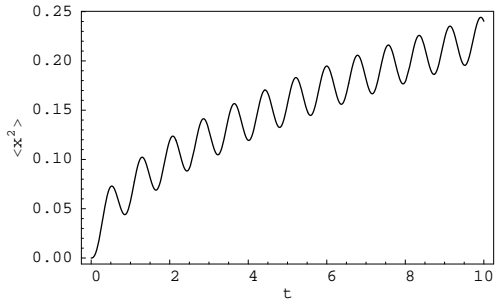}
\caption{$\eta =0$, $\omega_0=4$}\label{eta0bild}
\end{center} 
\end{figure}
In figure (\ref{eta0bild}) we see the MSD oscillating around the asymptotics $\sim t^{\frac{1}{2}}$. The bounding potential thus leads to subdiffusive behaviour.\\
The next special case is $\omega_0=0$, i.e. no external potential is present. In figure \ref{omega0bild} we see the from (\ref{msdeta0}) expected subdiffusive behaviour.\\
Let us next consider the case when the motion between the kicks is overdamped, i.e. $\omega'^2<0$. We see from figure
(\ref{overdamped}) an exponential decay of the MSD. The particle localizes at the attracting center.\\ 
Finally we consider the case $\omega'^2 >0$. This corresponds in the classical picture to damped
oscillating motion between the kicks.\\ 
\begin{figure}
\begin{center}
\includegraphics[height=4cm]{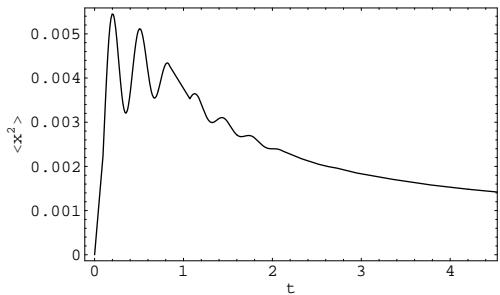}
\caption{Damped oscillations between the kicks, $\eta=1$, $\omega_0=10$}\label{oscillatingbild}
\end{center} 
\end{figure}
The behaviour of the  MSD is apparently similar to the behaviour of the classical position $x(t)$ of the Harmonic Oscillator. It is quite interesting however to see that for $\eta=0$ as well as for $\omega_0 =0$ the MSD shows subdiffusive behaviour, while when both effects are active the MSD exponentially decays and so localizes. In this context we want to mention that the long-time behaviour of the second-order moments for the Kramers equation proposed by Barkai et al. is calcuable even for $\gamma\neq 0$ and leads to loacalization even if $\eta=0$. As the Kramers equation of Metzler et al. describes subdiffusion, localization is here obtained too.\\
Summarizing the bounding force as well as the additional damping turns the superdiffusive-ballistic behaviour of the force-free case into the subdiffusive regime. If both effects are present the particle localizes eventually at the center of attraction.

\section{Conclusions}
We have investigated the generalized Kramers equation proposed in \cite{Eule} for a particle moving in a
a harmonic potential under the influence of damping. We presented the appropriate Kramers equation. For a 
fractional time evolution kernel we discussed limiting cases for the behaviour of the moments, i.e. $\omega_0=0$, $t\sim \infty$ and the general case for $\gamma=0$. For $\omega_0=0$ we obtained a transition from the superdiffusive-ballistic to the subdiffusive regime. The rate of convergence depends on the value of the additional damping.\\
For $\gamma=0$ we could obtain a closed expression for the mean square displacement. We presented numerical results for the mean square displacement and discussed qualitative features. In this case the harmonic potential as well as the additional damping shift the ballistic-superdiffusive behaviour, present in the force-free case, to the subdiffusive regime. For $\omega_0=0$ this also holds true for the general case. Interestingly only both effects together lead to a localisation of the particle.\\

\end{document}